\documentclass[aps,prl,amsmath,twocolumn,groupedaddress,showkeys,showpacs]{revtex4-1}
\usepackage{graphicx}
\usepackage{epsfig}
\newcommand{\beqy}{\begin{eqnarray}}
\newcommand{\eeqy}{\end{eqnarray}}

\begin{document}

\title{Crustal Entrainment and Pulsar Glitches}
\author{N. Chamel}
\affiliation{Institut d'Astronomie et d'Astrophysique, CP-226, Universit\'e Libre de Bruxelles, 
1050 Brussels, Belgium}

\begin{abstract}
Large pulsar frequency glitches are generally interpreted as sudden transfers of angular momentum between the neutron superfluid 
permeating the inner crust and the rest of the star. Despite the absence of viscous drag, the neutron superfluid is 
strongly coupled to the crust due to nondissipative entrainment effects. These effects are shown to severely limit the 
maximum amount of angular momentum that can possibly be transferred during glitches. In particular, it is found that the 
glitches observed in the Vela pulsar require an additional reservoir of angular momentum. 
\end{abstract}


\maketitle

{\it Introduction.} Since their fortuitous discovery by Jocelyn Bell and Anthony Hewish in 1967, more than 2000 pulsars
have been found~\cite{atnf}. Their identification as neutron stars~\cite{hae07}, the compact residues of type II supernova explosions
predicted by Baade and Zwicky in 1933~\cite{bz34}, was definitively established the next year after the discoveries of pulsars 
in the Crab and Vela supernova remnants. Pulsars are among the most accurate clocks in the Universe with periods ranging from 
1.4 milliseconds up to several seconds. The delays associated with the spin-down of the star are at most of a few  
of milliseconds per year. 

Nevertheless, irregularities have been detected in long-term pulsar timing observations~\cite{jod12}. In particular, some pulsars 
exhibit sudden increases in their rotational frequency $\Omega$. These ``glitches'', whose amplitude varies from 
$\Delta\Omega\slash \Omega\sim 10^{-9}$ up to $\sim 10^{-5}$ are generally followed by a relaxation over days to 
years and are sometimes accompanied by a sudden change of the spin-down rate from $\vert\Delta \dot\Omega\slash\dot\Omega\vert\sim 10^{-6}$ 
up to $\sim 10^{-2}$ (see, e.g., Sec. 12.4 in Ref.~\cite{lrr}). 

Soon after the first observations of glitches in the Vela and Crab pulsars, several scenarios were advanced~\cite{rud72}. 
In particular, glitches were thought to be the manifestations of starquakes, but this could not explain the frequent 
occurrence of Vela pulsar glitches~\cite{rud69}. A corequake model of Vela pulsar glitches was proposed~\cite{psr72}, but the 
existence of a solid core later appeared to be highly speculative (see, e.g., Ref.~\cite{hae07}). The long relaxation times following 
glitches provided strong evidence for the presence of superfluids in neutron star interiors and hinted at its possible role in the glitch mechanism itself~\cite{bay69,pac72}. 
Neutron-star superfluidity had been predicted and studied even before the discovery of pulsars~\cite{mig59,gk64}. Anderson 
and Itoh developed the fruitful idea that Vela like glitches are related to the dynamics of the neutron superfluid permeating 
the inner crust of neutron stars~\cite{and75}. 

{\it Vortex-mediated glitches.} The neutron superfluid is weakly coupled to the crust by mutual friction forces and thus follows 
its spin-down via a radial motion of quantized vortices away from the rotation axis unless vortices are pinned to the crust. In this 
case, the superfluid can rotate more rapidly than the crust. The lag between the superfluid and the crust induces a Magnus force 
acting on the vortices thereby producing a crustal stress. When the lag exceeds a critical threshold, the vortices are suddenly 
unpinned. As a result, the superfluid spins down and, by the conservation of the total angular momentum, the crust spins up leading 
to a glitch. This scenario found some support from laboratory experiments in superfluid helium~\cite{cam79,tsa80}. The good fit to 
the glitch data triggered further developments to explain the postglitch relaxation by the motion of vortices~\cite{alp81,alp85,jon93}.

In the meantime, it was argued that the core (supposed to contain superfluid neutrons and type I superconducting protons) is unlikely to 
play any role in glitch events~\cite{alp84} (see also Ref.~\cite{and06}). Due to nondissipative entrainment effects similar to 
those arising in superfluid $^3$He-$^4$He mixtures, neutron superfluid vortices carry a fractional magnetic quantum flux. Electron 
scattering off the magnetic field of the vortices leads to a (dissipative) mutual friction force acting on the superfluid. As a 
result, the core superfluid is strongly coupled to the crust and to the charged particles, thus following the long-term 
spin-down of the star caused by electromagnetic radiation. 

The confidence in the vortex-mediated glitch interpretation led to a new constraint on the structure of neutron stars hence also on 
the equation of state of dense matter~\cite{lnk99}. 
The latest models like the ``snowplow'' model~\cite{piz11} can reproduce various observations of pulsar glitches. 
However, many fundamental aspects of these models remain poorly understood. For instance, 
the strength of vortex pinning, which is one of the crucial microscopic inputs, has been a controversial issue over the past years 
(see, e.g., Sec. 8.3.5 of Ref.~\cite{lrr}). 
The mechanism that triggers the unpinning of vortices like superfluid instabilities~\cite{gla09} is also a matter of debate. 

More importantly, these models ignore the nondissipative entrainment 
effects in neutron-star crusts that have been shown to be very strong~\cite{cha05,cch05,cch05b,cch06,cha06,cha12}. In this Letter, the impact of crustal 
entrainment on pulsar glitches is studied. Using the latest pulsar glitch data~\cite{jod12}, it is shown that the neutron superfluid 
in the crust does not carry enough angular momentum to explain the Vela pulsar glitches. 

{\it Entrainment in neutron-star crusts.} 
It has been realized only recently that entrainment arises not only in the core of a neutron star but also in the crust
because unbound neutrons can be elastically scattered by the crustal lattice for specific wave vectors, as determined 
by Bragg's law~\cite{cha05,cch05,cch05b,cch06,cha06,cha12}. A neutron that is Bragg reflected cannot propagate and is therefore entrained by the crust. 
Unlike viscous drag, entrainment is nondissipative. Even if a neutron is not Bragg reflected, its
motion will still be affected by the crustal lattice. Neutron diffraction experiments are routinely performed to study 
crystal structures. The specificity of neutron-star crusts is that neutrons form a highly degenerate quantum liquid. 
Due to the Pauli exclusion principle, neutrons have different wave vectors and are therefore diffracted differently. 
Entrainment can be characterized by the density $n_n^{\rm c}$ of conduction neutrons, i.e. neutrons that are 
effectively ``free'', or equivalently by an effective neutron mass $m_n^\star=m_n n_n^{\rm f}/n_n^{\rm c}$ where 
$m_n$ is the bare neutron mass and $n_n^{\rm f}$ the density of unbound neutrons. Neutron conduction has been systematically 
studied in all regions of the inner crust using a state-of-the art crust model based on the band theory of 
solids~\cite{cha12}. Entrainment has been found to be very strong, especially in the intermediate region of 
the inner crust at average baryon densities $\bar n\sim 0.02-0.03$~fm$^{-3}$ where $n_n^{\rm c}\ll n_n^{\rm f}$ or 
equivalently $m_n^\star\gg m_n$.

{\it Pulsar glitch constraint.} Large pulsar glitches are usually interpreted as sudden transfers of angular momentum 
between the neutron superfluid in the crust and the rest of the star~\cite{alp85}. This model predicts that the ratio 
of their respective moments of inertia must obey the constraint~\cite{lnk99}
\beqy
\label{1}
\frac{I_{\rm s}}{I_{\rm c}}\geq \mathcal{G}\equiv A_g\frac{\Omega}{|\dot\Omega|}
\eeqy
where $A_g$ is the glitch activity parameter defined by the sum over glitches occurring during a time $t$
\beqy
\label{2}
A_g=\frac{1}{t}\sum_i\frac{\Delta\Omega_i}{\Omega}
\eeqy
while $\dot\Omega$ is the average spin-down rate. Both $A_g$ and $\dot\Omega$ can be measured from pulsar-timing observations. 
Since $I_{\rm s}\ll I_{\rm c}$, $I_{\rm c}$ can be replaced by the moment of inertia $I=I_{\rm s}+I_{\rm c}$ of the entire star. 
Approximating $I_{\rm s}$ by the moment of inertia $I_{\rm crust}$ of the crust, a constraint on the mass and radius of the Vela 
pulsar was derived in Ref.~\cite{lnk99}. This approximation treats all unbound neutrons as conducting ($n_n^{\rm c}=n_n^{\rm f}$), an 
assumption which turns out to be unrealistic~\cite{cha05,cha06,cha12}. Due to entrainment, the angular momentum $J_{\rm s}$ of the 
superfluid  depends not only on the angular velocity $\Omega_{\rm s}$ of the superfluid, but also on the observed angular velocity 
$\Omega$ of the star and can be expressed as~\cite{cc06}
\beqy
\label{3}
J_{\rm s}=I_{\rm ss} \Omega_{\rm s} + (I_{\rm s}-I_{\rm ss}) \Omega\, ,
\eeqy
with
\beqy
\label{4}
I_{\rm s}=\int m_n n_n^{\rm f} \varrho^2 {\rm d}^3r\, ,\hskip0.5cm I_{\rm ss}=\int m_n^\star n_n^{\rm f} \varrho^2 {\rm d}^3r\, ,
\eeqy
where $\varrho$ is the cylindrical radius. The constraint~(\ref{1}) thus becomes~\cite{cc06}
\beqy
\label{7}
\frac{(I_{\rm s})^2}{I I_{\rm ss}}\geq \mathcal{G}\, .
\eeqy
This inequality is much more stringent than (\ref{1}) because $I_{\rm ss}\gg I_{\rm s}$.

\begin{figure}
\includegraphics[scale=0.32]{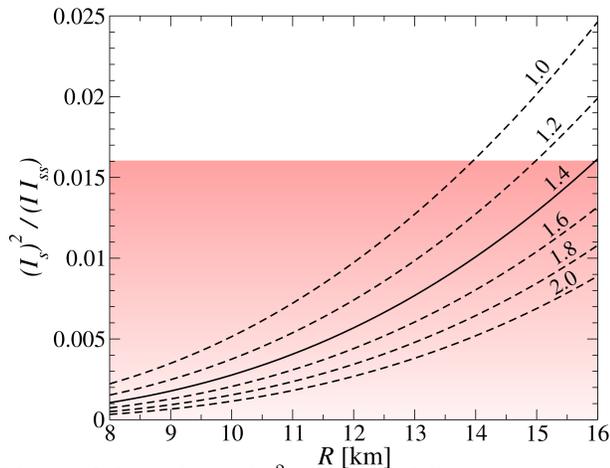}
\vskip -0.5cm
\caption{(Color online) $(I_{\rm s})^2/(I I_{\rm ss})$ for different neutron-star radii $R$ and masses $M$ from 
$1 M_\odot$ (upper curve) to $2 M_\odot$ (lower curve). The shaded area is excluded if Vela pulsar glitches
originate from the neutron superfluid in the inner crust with the crustal entrainment parameters of Ref.~\cite{cha12}. }
\label{glitch-cnt}
\end{figure}

{\it Results.} The ratio appearing in the left hand side of Eq.~(\ref{7}) can be decomposed as
\beqy
\label{8}
\frac{(I_{\rm s})^2}{I I_{\rm ss}}=\frac{I_{\rm crust}}{I_{\rm ss}}\left(\frac{I_{\rm s}}{I_{\rm crust}}\right)^2\frac{I_{\rm crust}}{I}\, .
\eeqy
In the thin crust approximation~\cite{lor93}, $I_{\rm ss}$ is given by
\beqy
\label{8b}
I_{\rm ss}\approx \frac{8\pi R^6}{3 G M}\left(1-\frac{2 G I}{R^3 c^2}\right)\int_{P_{\rm drip}}^{P_{\rm core}} \frac{n_n^{\rm f}(P) m_n^\star(P)}{\bar \rho(P)}\, {\rm d} P\, ,
\eeqy
where $M$ and $R$ are the neutron-star mass and radius,  $\bar \rho$ is the average mass density, $P_{\rm core}$ is the pressure at the crust-core transition and 
$P_{\rm drip} (\ll P_{\rm core})$  is the neutron-drip pressure. Corresponding expressions for $I_{\rm s}$ and $I_{\rm crust}$ are
obtained by replacing $n_n^{\rm f} m_n^\star$ in Eq.~(\ref{8b}) by $n_n^{\rm f} m_n$ and $\bar \rho$ respectively. Note that 
$I_{\rm ss}/I_{\rm crust}$ and $I_{\rm s}/I_{\rm crust}$ depend only on the crust properties, and can be written as 
\beqy
\label{9}
\frac{I_{\rm ss}}{I_{\rm crust}} \approx \frac{1}{P_{\rm core}}\int_{P_{\rm drip}}^{P_{\rm core}} \frac{n_n^{\rm f}(P)^2}{\bar n(P)n_n^{\rm c}(P)}\, {\rm d} P\, ,
\eeqy
\beqy
\label{10}
\frac{I_{\rm s}}{I_{\rm crust}} \approx \frac{1}{P_{\rm core}}\int_{P_{\rm drip}}^{P_{\rm core}} \frac{n_n^{\rm f}(P)}{\bar n(P)}\, {\rm d} P
\eeqy
where $\bar n= \bar\rho/m_n$ is the average baryon density. 
Integrating Eqs.~(\ref{9}) and (\ref{10}) with the trapezoidal rule using the results of \cite{cha12} summarized in Table~\ref{tab1}, 
we find $I_{\rm ss}\simeq 4.6 I_{\rm crust}$ and $I_{\rm s}\simeq 0.89 I_{\rm crust}$ leading to $(I_{\rm s})^2/I_{\rm ss}\simeq 0.17 I_{\rm crust}$. 
The ratio $I_{\rm crust}/I$ depends on the global structure of neutron stars. We have made use of Eq.~(47) of Ref.~\cite{lat00}. This formula 
was obtained by solving the equations of general relativity using a set of realistic dense-matter equations of state (EoS). 
Results for $(I_{\rm s})^2/(I I_{\rm ss})$ are shown in Fig~\ref{glitch-cnt}. Note that microscopic calculations based on 
chiral effective field theory~\cite{heb10} (and more generally any realistic EoS) as well as observations of 
x-ray binaries~\cite{ste10} indicate that neutron stars with $M=M_\odot$ have a radius $R\lesssim 13$~km.  

\begin{table}
\begin{tabular}{|c|c|c|c|}
\hline
$P$ (MeV fm$^{-3}$) & $n_n^{\rm f}/\bar n$ (\%) & $n_n^{\rm c}/n_n^{\rm f}$ (\%) \\
\hline
0.0004575 & 15.0 & 82.6 \\
0.0009886 & 61.1 & 27.3 \\
0.006097  & 82.6 & 17.5 \\
0.01507   & 86.0 & 15.5 \\
0.03820   & 87.9 & 7.37 \\
0.06824   & 89.1 & 7.33 \\
0.1068    & 86.6 & 10.6 \\
0.1561    & 89.1 & 30.0 \\
0.2183    & 89.2 & 45.9 \\
0.2930    & 89.4 & 64.6 \\
0.3678    & 100 & 64.8 \\
\hline
\end{tabular}
\caption{Entrainment parameters in the inner crust of cold nonaccreting neutron stars as obtained in \cite{cha12}.
$\bar n$ is the average baryon density, $n_n^{\rm f}$ is the density of unbound neutrons, and $n_n^{\rm c}$ is the density 
of conduction neutrons. The pressure $P$ was calculated using the formulae in Appendix B of \cite{pcgd12}.}
\label{tab1}
\end{table}

\begin{figure}
\includegraphics[scale=0.32]{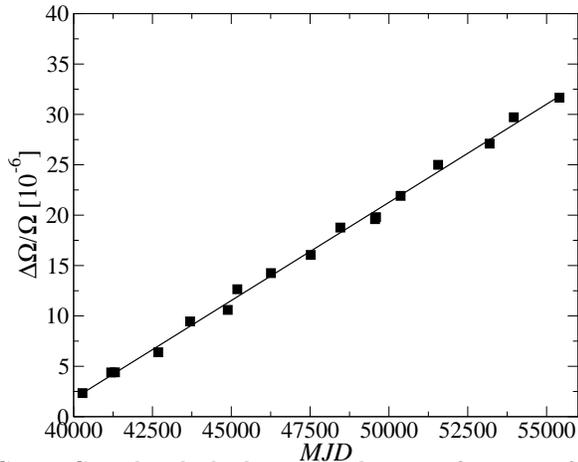}
\vskip -0.5cm
\caption{Cumulated glitch amplitudes as a function of the modified Julian date for the Vela pulsar from Ref.~\cite{jod12} (square symbols) and 
linear fit (solid line).}
\label{vela}
\end{figure}

\begin{figure}
\includegraphics[scale=0.32]{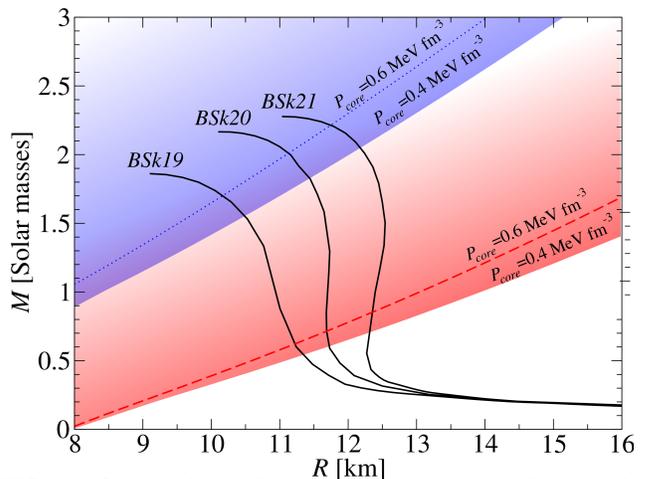}
\vskip -0.5cm
\caption{(Color online) Neutron-star mass-radius diagram for three different unified equations of state~\cite{cfpg11}.
The shaded area is excluded by Vela pulsar glitch data, assuming that only the neutron superfluid in the crust is
involved. The lower (upper) shaded area is the constraint obtained with (without) taking into account crustal entrainment. The 
sensitivity of these constraints with respect to the crust-core transition pressure $P_{\rm core}$ is indicated by 
the dashed line and the dotted line. The pressure $P_{\rm core}=0.4$~MeV~fm$^{-3}$ is the value found with the crustal 
model in Ref.~\cite{cha12}.}
\label{MR}
\end{figure}

Because it was the first observed pulsar to exhibit glitches, Vela has become the testing ground for glitch theories. Since 1969, 
17 glitches have been detected~\cite{jod12}. As shown in Fig.~\ref{vela}, the cumulated glitch amplitudes given by 
$\sum_i\Delta \Omega_i/\Omega=t A_g$ (with an appropriate choice of time origin) increases almost linearly with the time $t$. 
A linear fit yields $A_g\simeq 2.25\times 10^{-14}$~s$^{-1}$. 
With the angular frequency $\Omega=11.1946499395$~Hz and average spin-down rate $\dot\Omega=-1.5666\times 10^{-11}$~s$^{-2}$~\cite{atnf}, we find $\mathcal{G}\simeq 1.6\%$. A similar estimate has been obtained from a statistical analysis of Vela like pulsars~\cite{ly00}.
As illustrated in Fig.~\ref{glitch-cnt}, combining the glitch data with Eq.~(\ref{7}) leads to a constraint on the global neutron-star structure. 
This constraint, which can be approximately written as $R\geq 8.51+5.23 M$ or equivalently $M\leq 0.190 R -1.61$ with $M$ in $M_\odot$ and $R$ in km, 
is plotted in Fig.~\ref{MR}, together with three representative unified 
EoS spanning different degrees of stiffness of dense neutron matter, from the softest (BSk19) to the stiffest (BSk21), 
as obtained from microscopic calculations~\cite{cfpg11}. The sensitivity of the glitch constraint with respect to the corresponding 
crust-core transition pressure is also shown. This analysis implies 
that Vela (and more generally pulsars with Vela like glitches) should be less massive than our Sun ($M<0.6M_\odot$ for the softest 
EoS). Such a low mass neutron star is unlikely to be formed in a type II supernova explosion~\cite{stro01}. However, the 
association of the Vela pulsar with the eponymous supernova remnant is well established. 

{\it Discussion.} We are thus led to conclude that the neutron superfluid in neutron-star crusts does not 
carry enough angular momentum to explain large pulsar glitches like those observed in Vela, unless crustal 
entrainment and crust-core coupling are much weaker than considered here. A similar conclusion has been reached in Ref.~\cite{and12}.

The estimates of $m_n^\star$ obtained in \cite{cha12} agree closely with previous calculations~\cite{cha05} 
using a different model thus suggesting that strong crustal entrainment is generic. Moreover, $m_n^\star$ 
was found to be weakly dependent on the crystal structure~\cite{cha05}. The existence of nuclear ``pasta'' phases near the crust 
bottom (see, e.g., Sec. 3.3 of Ref.~\cite{lrr}), which we have ignored, might enhance the neutron conduction owing to the low dimensionality 
of such configurations. However, it has been argued that these pastas (if any) could only exist in a small region of the crust, at 
baryon densities above $\sim 0.06$~fm$^{-3}$, if the lowest-frequency quasiperiodic oscillation observed in giant flares from soft 
gamma-ray repeaters is to be interpreted as the fundamental torsional crustal mode~\cite{sot12}. Setting $n_n^{\rm c}=n_n^{\rm f}$ for 
$\bar n\geq 0.06$~fm$^{-3}$, the impact of pastas is found to be small since the ratio $I_{\rm ss}/I_{\rm crust}$ is reduced from 4.6 
to 4.3 whereas $(I_{\rm s})^2/(I_{\rm crust}I_{\rm ss})$ is raised from 0.17 to 0.19. In reality, $n_n^{\rm c}$ is never equal to 
$n_n^{\rm f}$ in any region of the crust, even in the presence of pastas~\cite{cha05,cch05}. On the other hand, the spin-orbit coupling 
(which was neglected in \cite{cha12}), would increase the number of entrained neutrons~\cite{cha05} and could be more important than 
pastas since it operates at all densities. We also anticipate that quantum and thermal fluctuations of ions about their equilibrium 
positions, crystal defects, impurities, and, more generally, any kind of disorder would presumably reduce (but not cancel) entrainment 
effects. Further work is needed to confirm these espectations. 

On the other hand, the strong coupling of the core to the crust that we have considered here relies on the assumption of type I 
superconductivity~\cite{alp84}. The observed rapid cooling of the neutron star in Cassiopeia A has recently provided 
strong evidence for core neutron superfluidity and proton superconductivity, but not on its type~\cite{pag11,sht11}. 
If the superconductor is of type II, the coupling time could be much longer~\cite{sed95b}. On the other hand, type II 
superconductivity has not only been argued to be incompatible with observations of long-period precession in 
pulsars~\cite{lnk03} but has also been questioned on theoretical grounds~\cite{chz07}. In fact, the superconductor 
might neither be of type I nor of type II~\cite{bab09}. In addition, neutron-star cores might contain other particle
species with various superfluid and superconducting phases. 

Removing the discrepancy between glitch models and observations thus requires a closer examination of crustal entrainment and 
crust-core coupling. The regularity of glitches illustrated in Fig.~\ref{vela} and the fact that ${\cal G}\lesssim 2\%$ suggest 
the existence of a reservoir of angular momentum in a limited region of the star, possibly in the outermost part of the core just below 
the crust (e.g., Refs.~\cite{sc99,per06}). This warrants further studies. 

This work also shed light on the importance of crustal entrainment, which has been generally overlooked even 
though they may have implications for other astrophysical phenomena such as quasiperiodic oscillations in soft gamma-ray repeaters~\cite{pas12}. 

\begin{acknowledgments}
The author thanks B. Link and A. Alpar for valuable discussions. This work was financially supported by FNRS (Belgium)
and CompStar, a Research Networking Programme of the European Science Foundation. The author thanks the Institute for 
Nuclear Theory at the University of Washington for its hospitality and the Department of Energy for partial support.
\end{acknowledgments}

\end{document}